\documentstyle[11pt,newpasp,twoside]{article}
\def\cent {$\omega$\thinspace Centauri}
\def\tuca {47\thinspace Tucanae}

\def\tuc {47\thinspace Tuc}

\def\arcmin              {$^{\prime}$} 
\def\arcm                {$^{\prime}$} 
\def\arcsec              {$^{\prime\prime}$} 
\def\arcs                {$^{\prime\prime}$}

\def\bily                {${ {10^9}}$yr}

\def\etal                {et\thinspace al.}

\def\kms                 {km\thinspace s$^{-1}$}

\def\mily                {${ {10^6}}$yr}

\def\msun                {$ M_{\odot}$}

\def\pmm                 {$\pm$}

\def\rh                  {$r_h$}

\def\edcomment#1{\iffalse\marginpar{\raggedright\sl#1\/}\else\relax\fi}
\reversemarginpar

\begin{document}
\title{Mass Segregation in Star Clusters} 

\author{Georges Meylan}

\affil{European Southern Observatory, Karl-Schwarzschild-Strasse 2, 
\hfill\break D-85748 Garching bei M\"unchen, Germany}

\begin{abstract}
Star clusters -- open  and globulars -- experience dynamical evolution
on time  scales shorter   than  their  age.   Consequently,  open  and
globular  clusters provide us  with  unique dynamical laboratories for
learning about    two-body    relaxation,  mass   segregation     from
equipartition of energy, and core collapse.  We review briefly, in the
framework of star clusters,  some elements related to  the theoretical
expectation of  mass  segregation, the  results from N-body  and other
computer   simulations, as   well    as  the now   substantial   clear
observational evidence.
\end{abstract}

\keywords{stars -- clusters -- stellar dynamics -- stellar photometry}


\section{Three Characteristic Time Scales}

The  dynamics   of any stellar  system   may be  characterized  by the
following three dynamical time scales: (i) the crossing time $t_{cr}$,
which is the time needed by a star to move across the system; (ii) the
two-body relaxation  time $t_{rlx}$, which is the  time  needed by the
stellar    encounters  to   redistribute   energies,  setting    up  a
near-maxwellian  velocity  distribution;   (iii)  the  evolution  time
$t_{ev}$, which is  the  time during which energy-changing  mechanisms
operate, stars   escape,  while the size   and profile  of  the system
change.

Several  (different and precise)  definitions exist for the relaxation
time.   The most   commonly  used  is the   half-mass relaxation  time
$t_{rh}$   of Spitzer  (1987,  Eq.~2-62),  where the   values for  the
mass-weighted mean square  velocity of the  stars and the mass density
are those evaluated at the half-mass radius of  the system (see Meylan
\& Heggie 1997 for a review).

In  the case  of globular  clusters,  $t_{cr}$ $\sim$ \mily, $t_{rlx}$
$\sim$ 100~\mily, and $t_{ev}$ $\sim$ 10~\bily.  Table~1 displays, for
open clusters,   globular  clusters,  and  galaxies,  some interesting
relations between the  above  three time  scales.  For  open clusters,
crossing time $t_{cr}$ and relaxation time $t_{rlx}$  are more or less
equivalent,  both being significantly  smaller than the evolution time
$t_{ev}$.  This means  that most open  clusters dissolve within a  few
gigayears.   For  galaxies,  the  relaxation  time  $t_{rlx}$  and the
evolution time   $t_{ev}$  are more or   less  equivalent,  both being
significantly larger than the crossing time $t_{cr}$.  This means that
galaxies are  not relaxed, i.e.,  not dynamically evolved.  It is only
for  globular clusters that  all three  time  scales are significantly
different, implying plenty of time  for a clear dynamical evolution in
these stellar systems,  although avoiding  quick evaporation  altering
open clusters.

Consequently, star    clusters -- open   and globular    --  represent
interesting  classes of   dynamical   stellar systems  in  which  some
dynamical processes take place on  time scales shorter than their age,
i.e., shorter than the Hubble time, providing us with unique dynamical
laboratories for learning  about two-body relaxation, mass segregation
from equipartition of energy, stellar collisions, stellar mergers, and
core  collapse.   All these  dynamical  phenomena  are related  to the
internal  dynamical evolution only,  and would also happen in isolated
glo\-bular clusters.   The external dynamical  disturbances ---  tidal
stripping by the  galactic gravitational  field --- influence  equally
strongly the dynamical evolution of globular clusters.

\begin{table}[t]
\caption{ Dynamical time scales for open clusters, globular clusters
and galaxies}
\begin{center}
\footnotesize
\begin{tabular}{|l|c|c|c|l|}
\hline  
 & & \\
open clusters &t$_{cr}$ $\sim$ t$_{rlx}$ $\ll$ t$_{ev}$ & quickly dissolved\\
 & & \\
globular clusters & t$_{cr}$ $\ll$  t$_{rlx}$ $\ll$  t$_{ev}$ & \\
 & & \\
galaxies          & t$_{cr}$ $\ll$  t$_{rlx}$ $\sim$ t$_{ev}$ &not relaxed\\
 & & \\
\hline
\end{tabular}
\end{center}
\end{table}


\section{Observed in all N-body Calculations}

Mass segregation was one of the early important results to emerge from
computer $N$-body   simulations  of star  clusters.    See,  e.g., von
Hoerner (1960) who made the first $N$-body calculations  with $N$ = 4,
8,  12,  and 16  bodies.  The  heavier  stars  would  gradually settle
towards  the center, increasing  their negative  binding energy, while
the lighter stars would preferentially populate the cluster halo, with
reduced binding energy. Later, direct integrations using many hundreds
of stars showed  the same tendency.   Soon  it was  also realized that
computation  of  individual  stellar  motions  could be   replaced  by
statistical methods.  The same mass segregation was observed in models
which  integrated the  Fokker-Planck   equation for many  thousands of
stars (e.g., Spitzer \& Shull 1975).

\begin{figure}
\plotfiddle{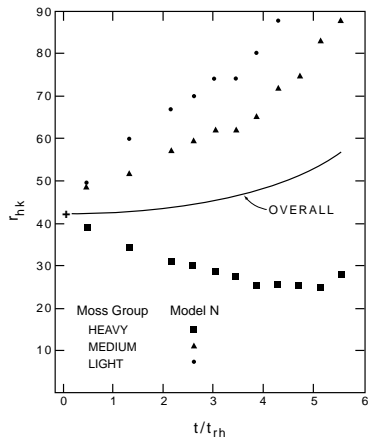}{6cm}{0}{45}{45}{-135}{-100}
\caption{Mass stratification: the  change,  as a function of  time, of
the  median radius \rh\ for  each  component (heavy, medium, and light
stars) of a three-subpopulation model,  as obtained from a Monte-Carlo
simulation by Spitzer \& Shull (1975).}
\label{}
\end{figure}

Mass segregation  is  expected from  the basic properties  of two-body
relaxation.   The time scale   for dynamical friction to significantly
decrease the  energy of a massive star  of mass $M$  is  less than the
relaxation time scale for lighter stars of mass  $m$ by a factor $m/M$
(see, e.g., Eq.~14.65 in Saslaw 1985).  As  massive stars in the outer
regions of a cluster lose energy to the lighter ones, they fall toward
the center and increase their  velocity. The massive stars continue to
lose the kinetic energy  they gain  by  falling and continue  to fall.
The lighter stars,  on the other  hand,  increase their  average total
energy and  move  into the  halo.   As light  stars rise   through the
system, their velocity decreases,  altering the local relaxation  time
for remaining massive stars.

Will this mass  segregation process have an  end, i.e. will the system
reach  an equilibrium?  Two   conditions would have  to be  satisfied:
mechanical equilibrium determined by the scalar virial theorem:

\begin{equation}
2 \langle T \rangle + \langle W \rangle = 0
\end{equation}

\noindent
and  thermal equilibrium determined   by equipartition of energy among
compo\-nents of different mass $m_i$:

\begin{equation}
m_i \langle v^2_i \rangle = 3kT
\end{equation}

All species  must have  the same  temperature,  so there is no  energy
exchange among the different species.


\section{Mass Segregation from Photometric Observations}

From a pure observational point of view, mass segregation has now been
observed  clearly in  quite a few  open  and globular clusters.  These
observational  constraints  are essentially  photometric:    different
families  of stars, located in  different areas of the color-magnitude
diagram    (CMD),  exhibit different   radial  cumulative distribution
functions.  Such an effect, when  observed between bi\-naries and main
sequence stars or   between  blue stragglers  and turn-off  stars,  is
generally interpreted   as an indication   of mass segregation between
subpopulations of stars with different individual masses.

We present hereafter examples  of observations of mass segregation  in
three different  kinds of star  clusters:  (i) in the  very young star
cluster R136, (ii) in a few open clusters, and (iii) in a few globular
clusters.

\subsection{In the Very Young Star Cluster R136}

The  Large Magellanic Cloud star cluster  NGC~2070  is embedded in the
30~Dora\-dus  nebula, the largest HII  region in  the Local Group (see
Meylan   1993 for a review).   The  physical size  of NGC~2070, with a
diameter  $\sim$ 40  pc, is typical  of  old  galactic and  Magellanic
globular clusters  and is also comparable to  the size of  its nearest
neighbor, the young globular cluster NGC~2100. With an age of $\sim$ 4
$\times$ 10$^6$~yr (Meylan 1993, Brandl \etal\ 1996), NGC~2070 appears
slightly younger than   NGC~2100 which  has  an age   of $\sim$  12-16
$\times$ 10$^6$~yr (Sagar \& Richtler 1991).

Brandl \etal\ (1996) obtained for R136, the  core of NGC~2070, near-IR
imaging in  $H,K$ bands with the ESO  adaptive optics system ADONIS at
the  ESO  3.6-m  telescope.   They  go  down to  $K$    = 20 mag  with
0.15\arcsec\ resolution  over  a   12.8\arcsec\ $\times$  12.8\arcsec\
field containing R136  off center.  They  present photometric data for
about 1000 individual stars of O, B, WR spectral  types.  There are no
red giants or supergiants in their field.

Brandl \etal\ (1996) estimate from their total  $K$ magnitude that the
total  stellar mass  within 20\arcs\  is  equal to  3 $\times$  10$^4$
\msun, with an upper limit on this  value equal to 1.5 $\times$ 10$^5$
\msun.  A   star  cluster with  a  mass  of this range   and a typical
velocity dispersion of $\sim$ 5 \kms\  would be gravitationally bound,
a conclusion not  immediately  applicable to  NGC~2070  because of the
important mass loss due to   stellar evolution experienced by a  large
number of its stars (see Kennicutt \& Chu 1988).

Mass segregation may have been observed in R136, the core of NGC~2070.
From their  luminosity function, Brandl  \etal\  (1996) determine, for
stars more massive than 12 \msun, a mean mass-function slope $x$ = 1.6
[$x$(Salpeter) = 1.35], but this value increases from $x$ = 1.3 in the
inner 0.4 pc to $x$ = 1.6 for 0.4 pc $<$ $r$ $<$ 0.8  pc, and to $x$ =
2.2 outside 0.8 pc.   The fraction of massive  stars is higher  in the
center of R136.  Brandl   \etal\ (1996) attribute these variations  to
the presence of  mass segregation.  Given the  very young age of  this
system, which may  still be experiencing  from violent relaxation, the
cause of this mass  segregation is not immediately  clear.  It  may be
due to a spatially  variable  initial mass  function, a  delayed  star
formation  in  the core, or  the  result of  dynamical  processes that
segregated an initially uniform stellar mass distribution.


\subsection{In Young and Old Open Clusters}

Obviously, the older   the cluster, the  clearer  the mass segregation
effect. One of the first  such clear cases  was observed by Mathieu \&
Latham (1986) in M67 which, with an age of about 5~Gyr,  is one of the
oldest galactic open   clusters.  They studied  the radial  cumulative
distribution  functions of  the  following  three  families of  stars:
single stars, binaries, and blue stragglers, the latter being possibly
the results of  stellar  mergers.  The radial  cumulative distribution
functions     of  binaries and      blue stragglers  are   similar and
significantly more concentrated than  the distribution function of the
single  stars.  In  such a  dynamically relaxed   stellar system, this
result may be explained  only  by mass  segregation between  stars  of
different individual masses.

\begin{figure}
\plotfiddle{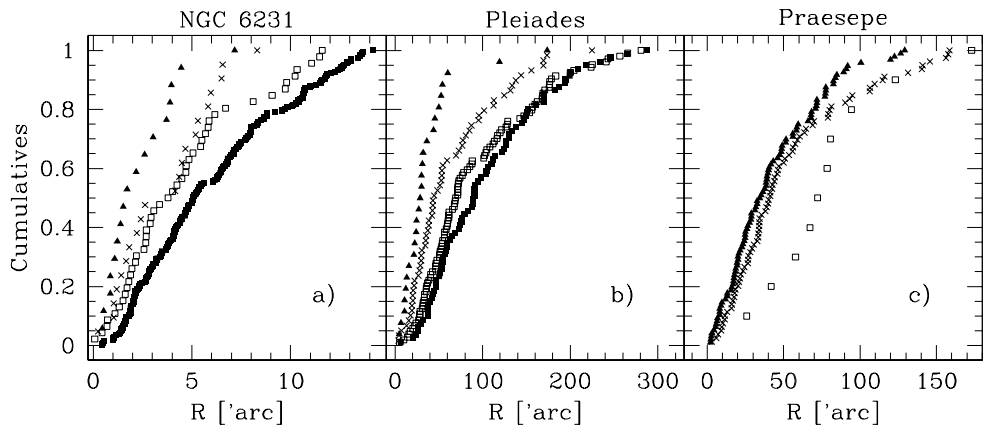}{6cm}{0}{70}{70}{-215}{-220}
\caption{Mass segregation  in  open clusters: cumulative distributions
of  stars  in identical relative  mass   intervals for  the three open
clusters  NGC~6231, the Pleiades, and  Praesepe, which have ages equal
to   4,   100, 800   Myr,     respectively.   Triangles for  $M   \geq
0.36~M_{max}$; crosses for  $0.23~M_{max} \leq M < 0.36~M_{max}$; open
squares for $0.14~M_{max} \leq M < 0.23~M_{max}$; filled squares for $
M < 0.14~M_{max}$.  From Raboud \& Mermilliod (1998).  }
\label{}
\end{figure}

In  one of the most recent  such studies,  Raboud \& Mermilliod (1998)
have observed some clear presence of  mass segregation (see Fig.~1) in
three open clusters -- NGC~6231, the  Pleiades, and Praesepe -- which,
with ages  equal to 4, 100,  800 Myr, respectively,  are significantly
younger than  M67.  The presence of mass  segregation in  the Pleiades
and Praesepe open  clusters is  expected   given the fact that   their
relaxation times are  shorter than their ages.  This  is not  the case
for  NGC~6231,  where the  presence  of mass   segregation may  be  as
problematic as it is in the case of R136.


\subsection{In Globular Clusters}

Because of their very high  stellar densities, globular clusters  have
been  hiding  for decades any  clear   observational evidence  of mass
segregation,  expected to  be   present essentially in   their crowded
central regions.  Differences in the  radial distributions of stars of
different luminosities/masses have  finally been de\-finitely observed
with  HST,  providing conclusive     observational evidence of    mass
segregation in the central parts of globular clusters.

One   of the most serious  and  detailed such  studies  is  the one by
Anderson (1997) who has used HST/FOC and HST/WFPC2 data to demonstrate
the presence  of mass   segregation in  the   cores of three  galactic
globular clusters: M92, \tuca, and \cent.

Anderson has first determined the luminosity  function of each cluster
at two  different locations in the  core.  Then he has  compared these
luminosity functions with those from King-Michie multi-mass models, in
the   cases with and  without  mass  segregation between the different
stellar species.

Fig.~3  displays the   comparison   between the  observed   luminosity
function (dots) and the model predictions  with (continuous lines) and
without (dashed lines) mass segregation, at the center of \tuca\ (left
panel) and at one core radius from the center  (right panel).  The two
different  models   differ strongly over a    large range in magnitude
(18-26 mag).  The observed  luminosity   function shows a  very  clear
agreement with  the model containing  mass segregation, and  rules out
completely any model without mass segregation (Anderson 1997).

\begin{figure}
\plotfiddle{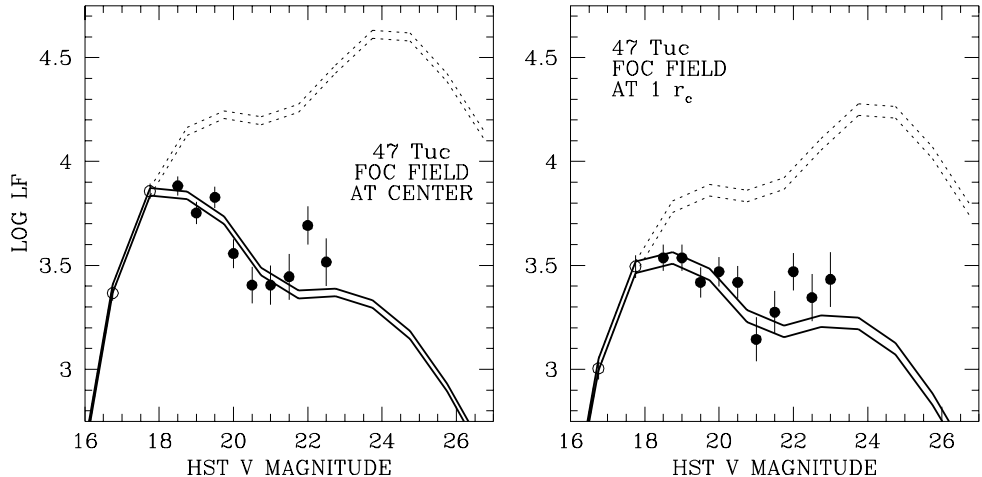}{6cm}{0}{70}{70}{-215}{-300}
\caption{Mass segregation  as observed in the  central parts of \tuca\
with HST/FOC data  (Anderson  1997): the observed luminosity  function
(dots)   agrees  with the    multi-mass   King-Michie model with  mass
segregation (continuous lines) and fails  totally to reproduce similar
models without mass segregation (dashed lines).}
\label{}
\end{figure}

The   globular  cluster M92 displays  results  very   similar to those
obtained for \tuca.   This is not  surprising given the fact that both
clusters have rather similar structural parameters and concentrations,
providing  similar central relaxation times of  the order  of 100 Myr.
This is not  the case for \cent,   which is the most  massive galactic
globular cluster and has a central relaxation time of about 6 Gyr.  As
expected, the  two model luminosity functions  (with  and without mass
segregation) computed  at the center  of  \cent\ differ only slightly,
and the observed  luminosity function is right  between the  curves of
the   two models.   The two  model   luminosity functions computed  at
16\arcmin\  from the center  (at  about 5 core   radii) do  not differ
significantly and consequently agree  similarly with the observations.
As  expected,    \cent, which has    had  hardly  any  time  to become
dynamically relaxed, even in its center,  exhibits a very small amount
of mass segregation (Anderson 1997).


\section{Mass Segregation Speeding Up the Dynamical Evolution Towards 
Core Collapse}

\subsection{Spitzer's Equipartition Instability}

As seen above, the various stellar species of a star cluster must have
the  same  temperature in  order    to have  equipartition of  energy.
Spitzer (1969) derived a criterion  for equipartition between stars of
two   different masses    $m_1$ and  $m_2$.     Let  us  consider  the
analytically tractable  case where the total  mass of the heavy stars,
$M_2$, is much smaller than the core mass of the system of the lighter
stars, $\rho_{c1} r_{c1}^3$, and  the individual heavy stars are  more
massive than   the light stars,   $m_2$ $>$  $m_1$.   In  such a case,
equipartition  will  cause the  heavy   stars (e.g.,  binaries  and/or
neutron stars) to form a small subsystem in the center  of the core of
the system formed by the light (e.g., main sequence) stars.

%
%

In equipartition, $m_2  \langle v_2^2   \rangle  = m_1  \langle  v_1^2
\rangle  =  3 m_1  \sigma^2$,  where $\sigma$   represents the central
one-dimensional dispersion of the light  stars.  It can be easily seen
(e.g., Binney \& Tremaine 1987) that equipartition cannot be satisfied
unless the following inequality holds:

%
%

\begin{equation}
{M_2 \over \rho_{c1} r_{c1}^3} \leq { 1.61 \over fg} \Bigl({m_1 \over
m_2}\Bigr)^{3/2}  
\end{equation}

\noindent
where $f$ and $g$ are  dimensionless constants.  When $M_2$ become too
large,  the  inequality is violated,    there  is the  ``equipartition
instability'' (Spitzer 1969), which has a simple physical explanation:
when the mass in   heavy  stars is  too  large,  these stars  form  an
independent high-temperature self-gravitating  system at the center of
the core of light stars.

In a realistic system with a distribution of stellar masses, the chief
effect of the equipartition instability  is to produce a dense central
core of  heavy stars, which contracts  independently from the  rest of
the   core.   However, as this   core  becomes denser  and denser, the
gravothermal instability dominates  over the equipartition instability
(Antonov 1962, Lynden-Bell \&  Wood 1968) and the cluster  experiences
core collapse (Makino 1996).

From  an  internal point  of view,  the   dynamical evolution of  star
clusters  is    driven by  two-body   relaxation,  mass   segregation,
equipartition instability, and  core collapse.  From an external point
of view, the  dynamical evolution of star   clusters is driven  by the
dynamical disturbances  due to the   crossing of  the galactic  plane,
which create tidal tails.  In whatever location, these stellar systems
are dynamically never at rest.

\subsection{Mass Segregation in M15, a prototypical core-collapse cluster}

The globular cluster  M15 has long been  considered as a prototype  of
the collapsed-core   star clusters.   High-resolution  imaging of  the
center of M15 has resolved the luminosity  cusp into essentially three
bright  stars.   Post-refurbishment Hubble  Space Telescope star-count
data confirm  that the 2.2\arcs\ core radius  observed by Lauer \etal\
(1991) and questioned by  Yanny \etal\ (1994),  is observed neither by
Guhathakurta \etal\ (1996) with HST/WFPC2  data  nor by Sosin \&  King
(1996, 1997) with HST/FOC data.   This surface-density profile clearly
continues to  climb steadily within  2\arcs.  It  is not  possible  to
distinguish at present  between a  pure power-law  profile and  a very
small   core (Sosin \&   King  1996, 1997).    Consequently, among the
galactic globular  clusters,  M15 displays one  of the  best  cases of
clusters caught in a state of deep core collapse.

Sosin \& King (1997) have estimated  the amount of mass segregation in
the   core of M15  from  their  HST/FOC  data: the  mass  functions at
20\arcs\  and 5\arcm\ from the   center clearly show substantial  mass
segregation for all stars with masses between 0.55 and 0.80 \msun.

\hskip -0.5truecm $\bullet$ the MF at $r$ = 20\arcs\ is best fit by a
power-law with slope $x$ = $-0.75$ \pmm\ 0.26,

\hskip -0.5truecm $\bullet$ the MF at $r$ = 5\arcm\  is best fit by a
power-law with slope $x$ = $+1.00$ \pmm\ 0.25.

These two slopes differ  at the 5-$\sigma$  level. Once  compared with
models, the amount of mass segregation is somewhat less than predicted
by a   King-Michie model, and  somewhat  greater  than  predicted by a
Fokker-Planck model. See also King \etal\ (1998) in the case NGC~6397.


\section{Mass Segregation from  Kinematical Observations}

Mass segregation  is also present in  kinematical  data, i.e.,  in the
radial velocities  and  proper motions of  individual stars.   So far,
radial velocities     have been obtained  essentially   only   for the
brightest stars, giants and subgiants, which have very similar masses.

It is only  recently that internal  proper motions of individual stars
have been obtained in globular clusters.   The following team (PI.  G.
Meylan, with CoIs.  D.  Minniti,  C.  Pryor, E.S.   Phinney, B.  Sams,
C.G.  Tinney, joined later  by J.  Anderson, I.R.  King,  and  W.  Van
Altena) have acquired HST/WFPC2 images in the core  of \tuca\ in three
different epochs (Oct.   1995 - Nov.   1997 -  Oct.  1999) defining  a
total time baseline of 4 years.  The choice  of the $U$ = FW300 filter
prevents saturation for  the  brightest stars and  allows simultaneous
measurement of proper motions  for the brightest  stars as well as for
stars more  than  two magnitudes  bellow the   turn-off (Meylan \etal\
1996).

For each   epoch  we have   15 images with   careful  dithering.  Each
measurement of the position of  a star has a  different bias since  in
each pointing the star is measured at a different pixel phase.  We use
an iterative  process on positions  and local  PSF determinations.  We
achieve a  position    accuracy of  0.020 pixel for    a single image,
amounting to 0.006 pixel for the  mean of 15 images.  This corresponds
to  0.3 mas  in the PC  frame and  0.6 mas  in  the WF2,  WF3, and WF4
frames, for about 14,000 stars in the core of \tuca\ (Anderson \& King
in preparation).

Preliminary results show a clear difference between the proper motions
of blue  stragglers and  stars of  similar magnitudes: the  former are
significantly  slower than  the  latter.   Since  blue stragglers  are
either binaries or mergers, with masses higher than the turn-off mass,
the above difference unveils the first kinematical observation of mass
segregation in a globular cluster (Meylan \etal\ in preparation).



\vskip 1.0truecm
{\bf Questions - Answers - Comments}

\vskip 0.5truecm

{\bf  Comment by  S. Portegies  Zwart} ~~From  a theoretical point  of
view, it is not always  clear what observers  considered as the center
of a star cluster and what theorists should use as the center. One can
use, for  example,  the geometric center,  the  area with  the highest
luminosity density, number density, mass density.

\vskip 0.3truecm

{\bf Comment by   G.M.}  ~~From an observational  point  of view,  the
determination  of the center  of a globular  cluster is also difficult
and  uncertain.   Ideally,  the  algorithm used should   determine the
barycenter of the stars, not of the light. In  the case of a collapsed
globular cluster like M15, which has a very small unresolved core, the
task is difficult because of the very small number of stars detectable
in such a small  area.  The uncertainty  in the position of the center
is of the order of the core radius value, i.e.,  about 0.2\arcsec.  In
the case   of a  globular   cluster like \tuca,   various methods give
positions  differing by 2-3   \arcsec.   Such a  large uncertainty  is
ne\-vertheless   acceptable, given  the   core radius  value of  about
25\arcsec.

\vskip 0.5truecm

{\bf  Question by H. Zinnecker}  ~~Is the mass segregation observed in
the 30  Doradus    cluster due  to  dynamical  evolution    or due  to
preferential birth of the more massive  stars near the cluster center?
Can we distinguish between these two possibilities?

\vskip 0.3truecm

{\bf Answer by G.M.} ~~It is not known if  the mass segregation is the
consequence of dynamical evolution or of a  flatter IMF in the center.
I fail  to see  any  reliable way   to distinguish between  these  two
possibilities.

\vskip 0.5truecm

{\bf Question by P. Kroupa} ~~The globular  clusters \tuca\ and \cent\
do not appear to have a pronounced  binary sequence in color-magnitude
diagrams,     whereas  other     globulars   have pronounced    binary
sequences. Does this imply different dynamical histories?

\vskip 0.3truecm

{\bf  Answer by  G.M.}  ~~It would   be   interesting to  compare  the
locations  where these various    color-magnitude diagrams have   been
obtained.  In the case of  \tuca, the excellent photometry we obtained
is  for stars  right in  the center,  where  encounters and collisions
operate and  probably decrease the  fraction of binaries.  It would be
interesting to make a precise  comparative study,  for a few  globular
clusters, for which  we would have data  from the  same instrument and
reduced with  the  same   software, in fields   at the   same relative
distance from the center.

\vskip 0.5truecm

{\bf Question by D. Calzetti}  ~~About studies which find flatter IMFs
in the centers of clusters: do you think that these studies may suffer
from  effects of crowding towards  the  cluster center, and therefore,
find a flatter IMF because of this?

\vskip 0.3truecm

{\bf Answer by G.M.} ~~Yes, definitely!  Crowding is always present in
photometry of globular clusters, especially  in observations from  the
ground.  Ne\-vertheless, I think that some careful photometric studies
using HST data have provided reliable results in relation to IMF slope
(see, e.g., King \etal, 1998, ApJ, 492, L37).

\vskip 0.5truecm

{\bf Question by C. Boily} ~~Core motion: has the relative position of
core to  outer    envelope been studied  for   candidate core-collapse
clusters, e.g., \tuc?

\vskip 0.3truecm

{\bf Answer by G.M.}

It would be an interesting  study, which, as  far as  I know, has  not
been done yet.   It is partly due to  the  difficulties in determining
precisely  the  center  of  the    core  and to the  difficulties   in
determining precisely outer isophotes which  suffer from very low star
counts and are strongly  polluted  by foreground stars and  background
galaxies.

\end{document}